\documentclass[slac_one]{revtex4}
\usepackage{graphicx}
\usepackage{fancyhdr}
\usepackage{amssymb}
\usepackage{lscape,amsmath}
\usepackage{latexsym}
\begin{document}

\title{{\small{2005 International Linear Collider Workshop - Stanford,
U.S.A.}}\\ 
\vspace{12pt}
Impact of tau polarization on the study of the MSSM charged Higgs
bosons in top quark decays at the ILC}

\author{E. Boos}
\affiliation{Skobeltsyn Institute of Nuclear Physics, MSU, 
119992 Moscow, Russia}
\author{V. Bunichev}
\affiliation{Skobeltsyn Institute of Nuclear Physics, MSU, 
119992 Moscow, Russia}
\author{M. Carena}
\affiliation{Fermi National Accelerator Laboratory, Batavia, IL 60510, USA}
\author{C.E.M. Wagner}
\affiliation{High Energy Physics Division, 
Argonne National Laboratory, Argonne,
IL 60637, USA}
\affiliation{Enrico Fermi Institute, Univ. of Chicago,
5640 S. Ellis Ave., Chicago, IL. 60637, USA}

\begin{abstract}
The process of top quark pair production at the ILC with 
subsequent decays of one
of the top quarks to a charged Higgs boson and $b$-quark is considered.  
The charged Higgs decays to tau leptons whose polarization is the
opposite to those coming from $W$~bosons.  This difference is
reflected in the energy distributions of the tau decay products in the
top quark rest frame, which can be reconstructed at the ILC using the
recoil mass technique.  We present an analysis including spin
correlations, backgrounds, ISR/FSR and beamstrahlung, and show that a fit 
of the shape of the pion energy spectrum yields the charged Higgs
boson mass with an accuracy of about 1~GeV.
\vspace*{0.6cm}\\
Pre-print numbers: ANL-HEP-CP-05-60, FERMILAB-CONF-05-265-T, EFI-05-06.
\end{abstract}

\maketitle

\thispagestyle{fancy}


\section{CHARGED HIGGS AND TAU POLARIZATION} 
The mechanism of spontaneous electroweak symmetry breaking in the MSSM leads to  
a Higgs sector with five physical states: two CP-even Higgs bosons, 
$h$ and $H$, one CP-odd one, $A$, and two charged Higgs bosons $H^{\pm}$
(If CP is violated then the three neutral Higgs bosons will not
have definite CP-parity.).
The discovery of the charged Higgs bosons would be very important,
as it would show directly that the Higgs sector has more complicated 
structure than the one in the Standard Model, thus providing clear
evidence for physics beyond the SM.

The charged Higgs boson couples strongly to the fermions of the 
third generation. If the charged Higgs is heavy
($M_{H^{\pm}}>M_t$), it can easily be detected in 
decays to top and bottom quarks ($H^- \to \bar{t}b$) 
by examining the $\bar{t}b$ invariant mass spectrum.
In this case, $M_{H^\pm}$ may be measured
to, at best, about $5$~GeV at the LHC~\cite{Assamagan:2004tt}, 
and about $1$~GeV at an $e^+e^-$ linear collider running at
an energy of $800$~GeV and collecting a luminosity of
$500$~fb$^{-1}$~\cite{Aguilar-Saavedra:2001rg}. 
In a number of other scenarios, however, the charged Higgs bosons 
will be rather light ($M_{H^{\pm}}<M_t$).  A precise measurement 
of $M_{H^{\pm}}$ in that range is a challenging task for any collider.  
In this case the charged Higgs decays dominantly to a tau lepton and neutrino 
($H^{\pm} \to \tau^{\pm} \nu$), and it is impossible to reconstruct directly 
the invariant mass of the di-tau final state.  However, due to the
polarization of the $\tau^{\pm}$ leptons, the energy of the $\tau^{\pm}$
decay products depends strongly on $M_{H^{\pm}}$, a feature that 
can be exploited to extract $M_{H^{\pm}}$ indirectly.  
The main background to a $H^{\pm}$ signal in the $\tau^{\pm}$ decay 
mode comes from the $W$-boson decays $W^\pm \to \tau^\pm \nu$. 
However, thanks to the different structure of the $H^{\pm}$ and 
$W^{\pm}$ electroweak interactions to $\tau^{\pm}$-leptons, 
the tau polarization is very different which will allow a separation
of $H^{\pm}\to\tau^{\pm}\nu$ and $W^{\pm}\to\tau^{\pm}\nu$
on a statistical basis. More specifically, the $\tau^{\pm}$ decay products
($\tau^+ \to \pi^+ \bar{\nu}$, or $\tau^+ \to \rho^+ \bar{\nu}$, etc.) 
have strikingly different topologies according as to whether they 
originate from a parent $W^{\pm}$ or $H^{\pm}$.
The importance of tau polarization in searches for charged Higgs bosons 
has already been stressed in previous studies~\cite{polarization}.

\begin{figure*}[t]
\centering
\includegraphics[width=75mm]{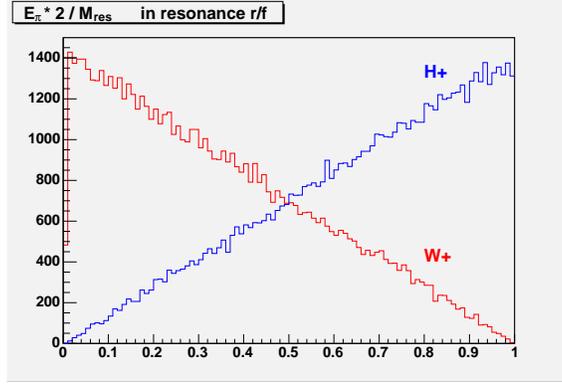}
\caption{$\pi^{\pm}$ meson energy spectrum in the resonance rest frame.} \label{res_rest}
\end{figure*}

The key point is that $\tau^-$ leptons arising from 
$H^- \to \tau^- \nu$ decays are almost purely right-handed, 
in contrast to the left-handed $\tau^-$ leptons which arise from $W^-$ 
decays. This contrast follows from the helicity flip nature of 
the Yukawa couplings of the Higgs fields, and the helicity 
conserving nature of the gauge interactions.
The most dramatic difference is seen in the energy distribution for
the single pion channel ($H^+/W^+ \to \tau^+ \nu \to \pi^+ \nu \bar{\nu}$)
in the rest frame of the parent boson ($W^{\pm}$ or $H^{\pm}$) --
see Fig.~\ref{res_rest}. 

In practice, it is nearly impossible to reconstruct the rest frame of the
$W^+$ or $H^+$ bosons because the momentum of the neutrinos cannot be
measured. Instead, one can use the top quark pair production channel
with top quark decays ($t \to H^+ b$) and reconstruct the top quark
rest frame. This is possible at a linear collider when the second produced top
quark decays hadronicly, in which case one uses the recoil mass 
technique analogous to those in Higgs boson studies~\cite{Aguilar-Saavedra:2001rg, recoiltechnique}.

The distributions for the energy of the single pion in the top rest
frame has the following form~\cite{sfermion} for $H^{\pm}$ and
$W^{\pm}$ cases, respectively.
\begin{eqnarray}\label{eq:formula}
\frac{1}{\Gamma}\frac{d\Gamma}{dy_{\pi}}=\frac{1}{x_{max}-x_{min}}
\begin{cases}
(1-P_{\tau})log\frac{x_{max}}{x_{min}} +
2P_{\tau}y_{\pi}(\frac{1}{x_{min}}-\frac{1}{x_{max}}),
&\text{$0<y_{\pi}<x_{min}$}\\
(1-P_{\tau})log\frac{x_{max}}{y_{\pi}} +
2P_{\tau}(1-\frac{y_{\pi}}{x_{max}}),
&\text{$x_{min}<y_{\pi}$}
\end{cases}
\end{eqnarray}
where $y_{\pi}=\frac{2E_{\pi}^{top}}{M_{top}}$,
$x_{min}=\frac{2E_{\tau}^{min}}{M_{top}}$, 
$x_{max}=\frac{2E_{\tau}^{max}}{M_{top}}$, 
$E_{\tau}^{min}=\frac{M_{R}^2}{2M_{top}}$,
$E_{\tau}^{max}=\frac{M_{top}}{2}$.
For the $W$~boson, $P_{\tau} =-1$, and for the charged Higgs boson,
$P_{\tau} = 1$. 
\footnote{For simplicity, in Eq.~\ref{eq:formula} we have neglected the b-quark and tau-lepton
masses, while these masses have been included in all numerical
simulations.}

\begin{figure*}[t]
\centering
\includegraphics[width=75mm]{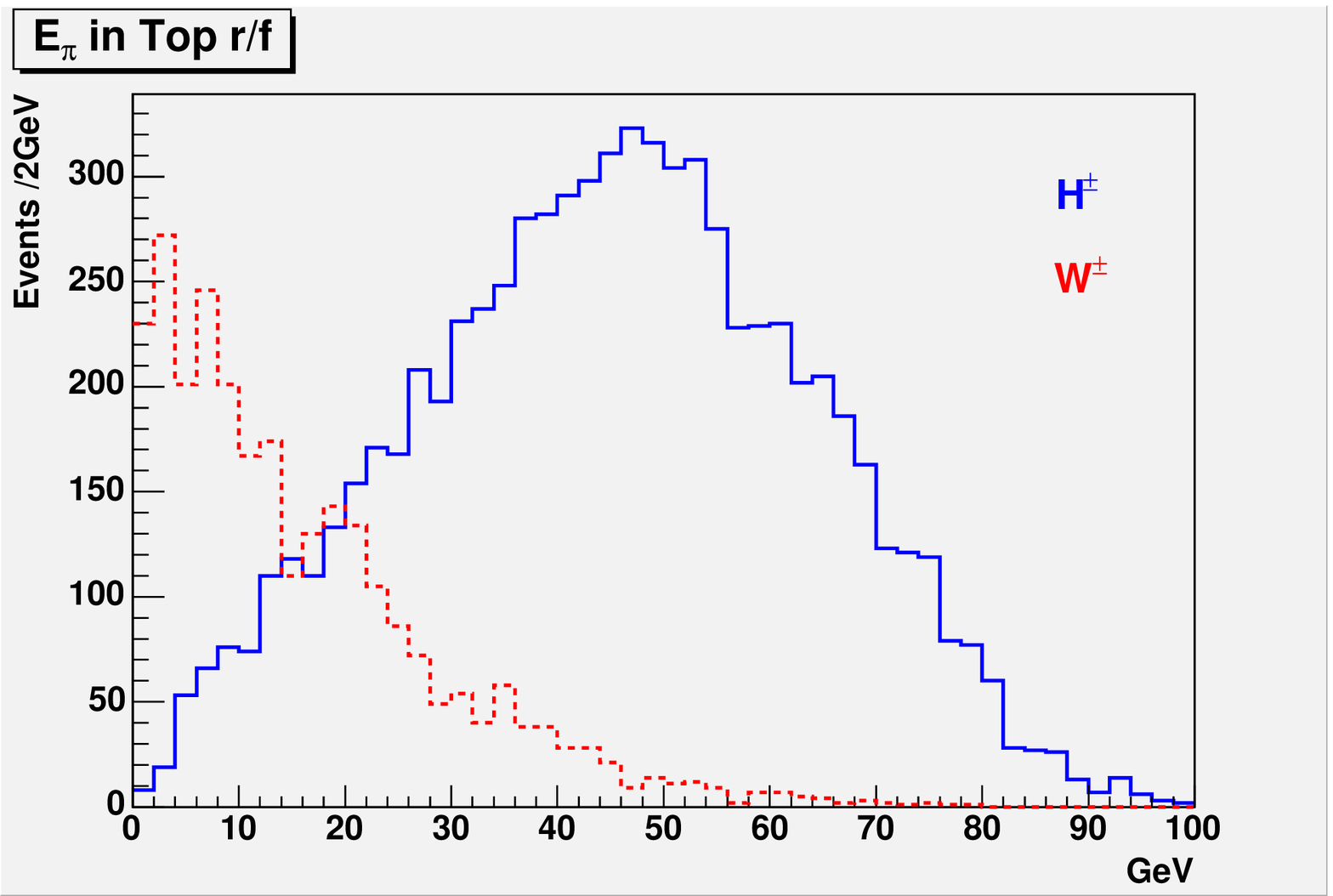}
\includegraphics[width=75mm]{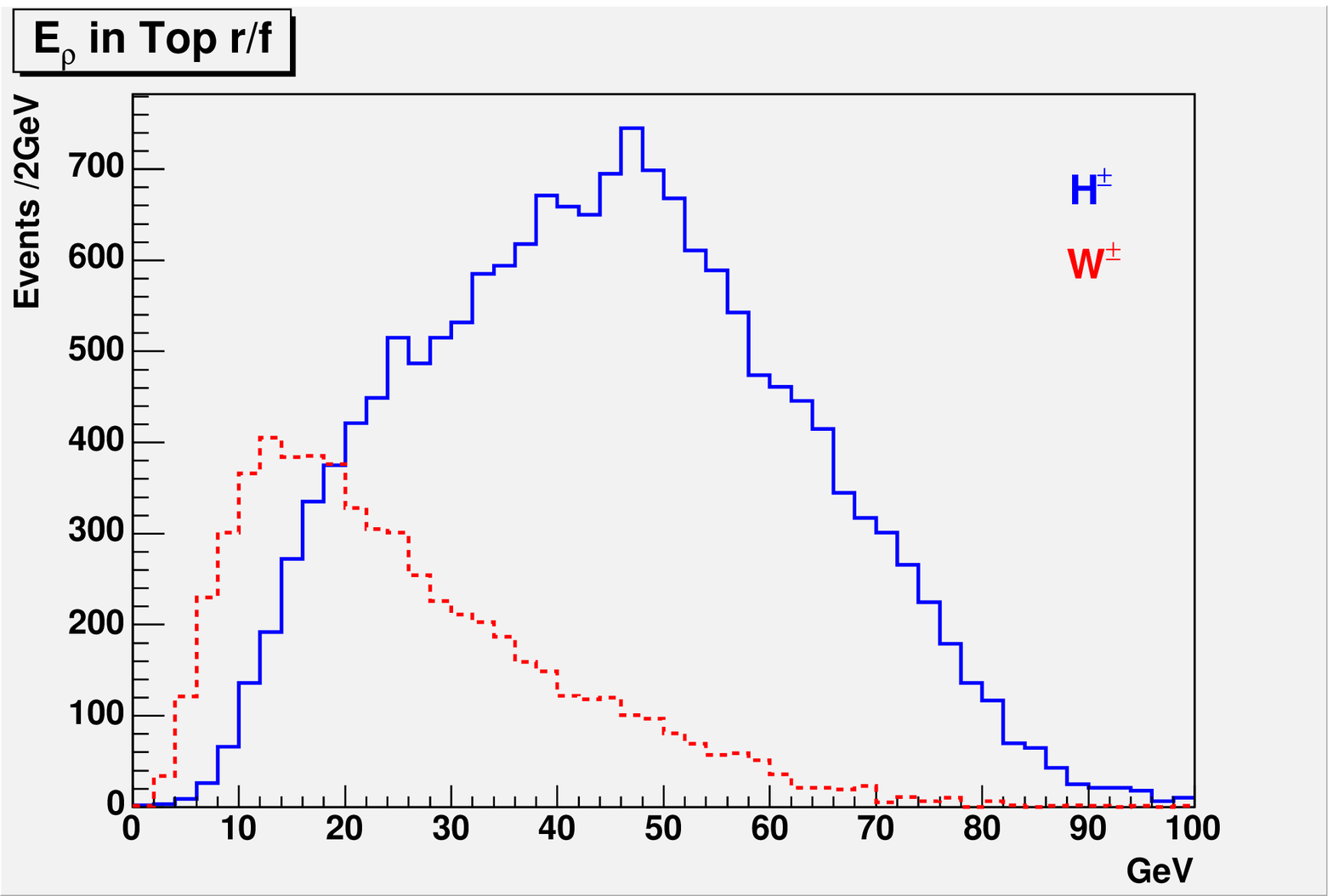}
\caption{The energy spectrum of the $\pi^{\pm}$ meson (left) and 
$\rho^{\pm}$ meson (right). The dotted line corresponds to the
background, and the solid one to signal. \label{Pi_top}} 
\end{figure*}

The energy distribution for a pion from $H^{\pm}$ decay has a maximum at the point 
$E(\pi^{\pm})={M_{H^{\pm}}}^2/(2M_t)$, as shown in Fig.~\ref{Pi_top}~(left). 
This dependence allows one to extract $M_{H^{\pm}}$ from the shape of
the spectrum. One can also take into account $\rho^{\pm}$ decay channel 
($\tau^+ \to \rho^+ \nu$), which has twice the branching ratio as the
pion channel.  The shape of the $\rho^{\pm}$ energy distribution is
more complicated, however, and is less sensitive to $M_{H^{\pm}}$,
as shown in Fig.~\ref{Pi_top}~(right). More detailed studies are
needed for this channel.

\section{EFFECTIVE {\boldmath${\bar{t}bH^{\pm}}$} INTERACTION IN THE MSSM}
For our simulations described in the next section, we use the
effective Lagrangian approach for the MSSM presented in
Ref.~\cite{lagrangian}.  
The effective Lagrangian of charged Higgs interaction with 
fermions of the third generation has the form:
\begin{equation}
 L \simeq \frac{g}{\sqrt2 M_W} \frac{\bar{m}_b(Q) \tan\beta}
{1+\Delta m_b} [V_{tb}H^+\bar{t}_L b_R(Q)+h.c.],
\end{equation}
where $\bar{m}_b$ is the running bottom mass in the $\overline{
\mathrm{MS}}$ scheme.
The width of the top quark decay to a charged Higgs and b-quark 
takes the form:
\begin{equation}
\label{GammaMSSM}
\Gamma_{MSSM}(t \to b H^+) \simeq \frac{\Gamma_{QCD}^{imp}(t \to b H^+) }
{(1+\Delta m_b)^2}
\end{equation}
where
\begin{equation}
\Delta m_b=\frac{\Delta h_b}{h_b}\tan\beta \sim \frac{2\alpha_S}{3\pi} 
\frac{\mu M_{\tilde g}}{max(m_{\tilde b_1}^2,m_{\tilde b_2}^2,M_{\tilde g}^2)}\tan\beta+\Delta_b^{\tilde t \tilde\chi^+} 
\end{equation}
\begin{equation}
\Delta_b^{\tilde t \tilde\chi^+} \sim \frac{h_t^2}{16\pi^2} 
\frac{\mu A_t}{max(m_{\tilde t_1}^2,m_{\tilde t_2}^2,\mu^2)}\tan\beta. 
\end{equation}
The $\Delta m_b$ correction
proceeds from the one-loop vertex corrections, which modify the relation
of the bottom Yukawa coupling to the bottom quark mass. Similar
corrections are induced in the neutral Higgs sector~\cite{Mrenna}.
The QCD improved top-quark decay width 
is given by~\cite{lagrangian},
\begin{eqnarray}
\Gamma_{QCD}^{imp}(t \to bH^+)=\frac{g^2}{64\pi M_W^2}m_t(1-q_{H^+})^2 \bar{m_b}^2(m_t^2) \tan^2\beta \times\nonumber\\
\left\{ 1+\frac{\alpha_S(m_t^2)}{\pi} 
\left[ 7-\frac{8\pi^2}{9}-2\log(1-q_{H^+})+
2(1-q_{H^+})\right.\right.\nonumber\\
\left.\left.+\left(\frac{4}{9}+\frac{2}{3}\log(1-q_{H^+})\right)(1-q_{H^+})^2 
\right] 
\right\} 
\end{eqnarray}
The above expression, Eq.~(\ref{GammaMSSM}), 
takes into account the dominant 
supersymmetric  corrections to all orders in perturbation theory.
It also includes the resummation of the large 
$m_t/m_b$ and $m_H^{\pm}/m_b$ logarithms, which amounts to replacing $m_b$ 
by the running bottom mass at the proper scale~\cite{lagrangian}, as
well as all finite one-loop QCD corrections.
 
In the numerical similations, 
for any  given set of  supersymmetric particle mass parameters, we
have computed the $\Delta m_b$ corrected charged Higgs Yukawa couplings 
with fermions of the third generation, by means of
the program CPSuperH~\cite{CPsuperH}.

\section{SIMULATIONS}

The numerical simulations have been performed assuming a center of mass
energy $\sqrt{s} = 500$~GeV and a total integrated luminosity
${\cal{L}} = 500$~fb$^{-1}$.
We have performed detailed computations and Monte Carlo simulations for
three different sets of MSSM parameters, all leading to light charged
Higgs bosons. The first two scenarios are based on the mass parameters
$M_Q = M_U = M_D = 1$~TeV, $M_{\widetilde{g}} = M_2 = 1$~TeV,
$A_t = 500$, $\mu = \pm 500$~GeV, and $\tan\beta = 50$, which give
$M_{H^{\pm}} = 130$~GeV.  The branching ratio $BR(t\to H^\pm b)$
will be enhanced or suppressed depending on the sign of~$\mu$.
The third parameter set falls in the so-called ``intense-coupling 
regime'' where the neutral Higgs boson masses are close to each 
other, and they couple strongly to fermions of the third 
generation~\cite{intense}.  In this case, we have $\mu = 1000$~GeV,
$\tan\beta = 30$ and $M_{H^{\pm}} = 146$~GeV.  The branching ratios
for these three examples are
\begin{center}
\begin{tabular}{crrl}
{\em (i)}   & $M_{H^\pm} = 130$~GeV, &
 \hspace*{10pt} $\mu < 0$ and $\tan\beta=50$: \hspace*{20pt}& 
  $BR(t\to H^+ b) = 0.24$ \cr
{\em (ii)}  & $M_{H^\pm} = 130$~GeV, &
 \hspace*{10pt} $\mu > 0$ and $\tan\beta=50$: \hspace*{20pt}& 
  $BR(t\to H^+ b) = 0.091$ \cr
{\em (iii)} & $M_{H^\pm} = 146$~GeV, &
 \hspace*{10pt} $\mu > 0$ and $\tan\beta=30$: \hspace*{20pt}& 
  $BR(t\to H^+ b) = 0.063$ \cr
\end{tabular}
\end{center}
These points are not excluded by Tevatron searches
which lead to the bound $BR(t\to H^+b)<0.42$ at 95\%~C.L.
when $M_{H^{\pm}} < 150$~GeV~\cite{TEVBOUND}.
The couplings have been implemented in CompHEP~\cite{comphep}, 
which has been used to compute the cross sections for signal and
background processes, including decays of top to $W^\pm$ and $H^\pm$
which subsequently decay to polarized $\tau^\pm$~leptons.
CompHEP was also used to generate events, and effects from initial
state radiation and Beamstrahlung were included. Polarized $\tau^\pm$ 
decays have been simulated using TAUOLA~\cite{tauola} interfaced to 
CompHEP. Hadronization and energy smearing in the final state are 
accounted for by means of PYTHIA~\cite{PYTHIA} using the
CompHEP-PYTHIA~\cite{interface} interface based on Les Houches 
Accord~\cite{Boos:2001cv}. Effects from final state radiation
have been implemented using the PHOTOS library~\cite{photos}.

A brief description of our fitting procedure is given here.
The fitting function is the sum of two functions, one for signal and
one for background.  For simplicity and without loss in accuracy we 
chose the following form for the charged Higgs signal piece,
which is motivated by the theoretical spectrum:
\begin{eqnarray}
H(x)= par_1 k_0 (|x-k_1|+k_2 x-k_3),
&&\text{$x=E_\pi$}
\end{eqnarray}
where:
$k_0={M_t^3}/{[par_2^2 (par_2^2-M_t^2)]}$,
$k_1={par_2^2}/{(2M_t)}$,
$k_2={2 par_2^2}/{M_t^2}-1$,
$k_3={par_2^2}/{(2 M_t)}$, with $par_2=M_{H^{\pm}}$.
There are two free parameters: the parameter $par_1$ gives the 
overall normalization of the signal events, and the parameter $par_2$
is the charged Higgs mass itself.  The function $H(x)$ has zeros at the
points $E_\pi=0$ and $E_\pi=M_t/2$,  and a maximum at the 
point $E_\pi=M_{H^{\pm}}^2/{(2 M_t)}$ as follows from the 
theoretical spectrum.  
For the background function we use 
a forth-order polynomial of the following form:
\begin{eqnarray}
B(x) = par_3 f(x),
&&\text{$x=E_\pi$}
\end{eqnarray}
where  $f(x)=c_0+c_1 x+c_2 x^2+c_3 x^3+c_4 x^4$,
and the fitting parameter $par_3$ gives the normalization of the
number of the background events. The coefficients  $c_0,\dots,c_4$ 
are fitted from the shape the background pion energy distribution.

We use the method of maximum likelihood to fit a spectrum created
from simulated signal and background events, and obtained the
following results for the three cases described above
\footnote{We are taking into account the expected accuracy, of about 100~MeV,
for the top quark mass determination from the top-quark pair threshold
scan at the ILC (see Ref. \cite{Aguilar-Saavedra:2001rg}), which adds only a small
contribution to the charged Higgs mass uncertainties.}

\begin{figure*}[t]
\centering
\includegraphics[width=75mm]{Pi_top_500gv.eps}
\includegraphics[width=75mm]{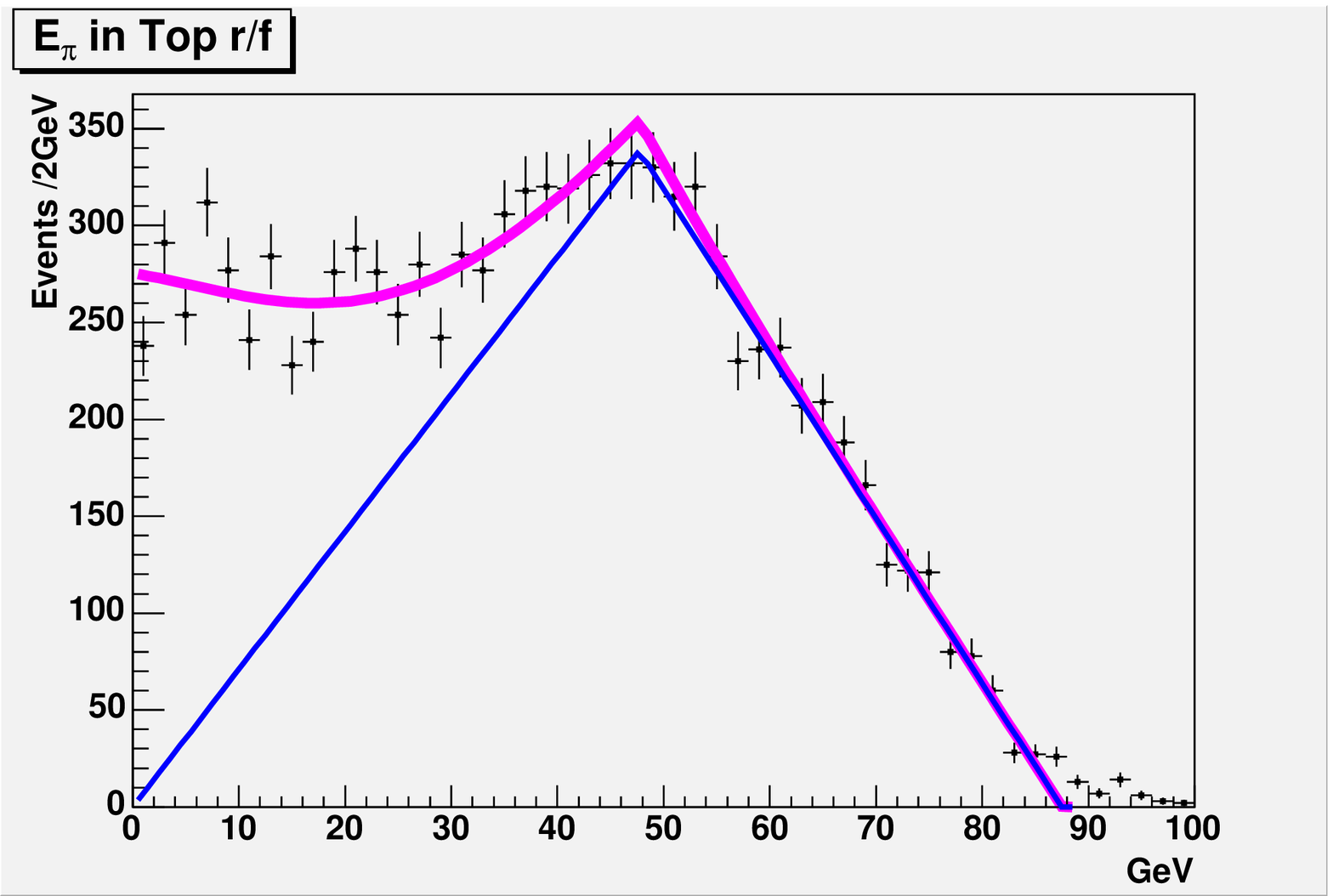}
\includegraphics[width=75mm]{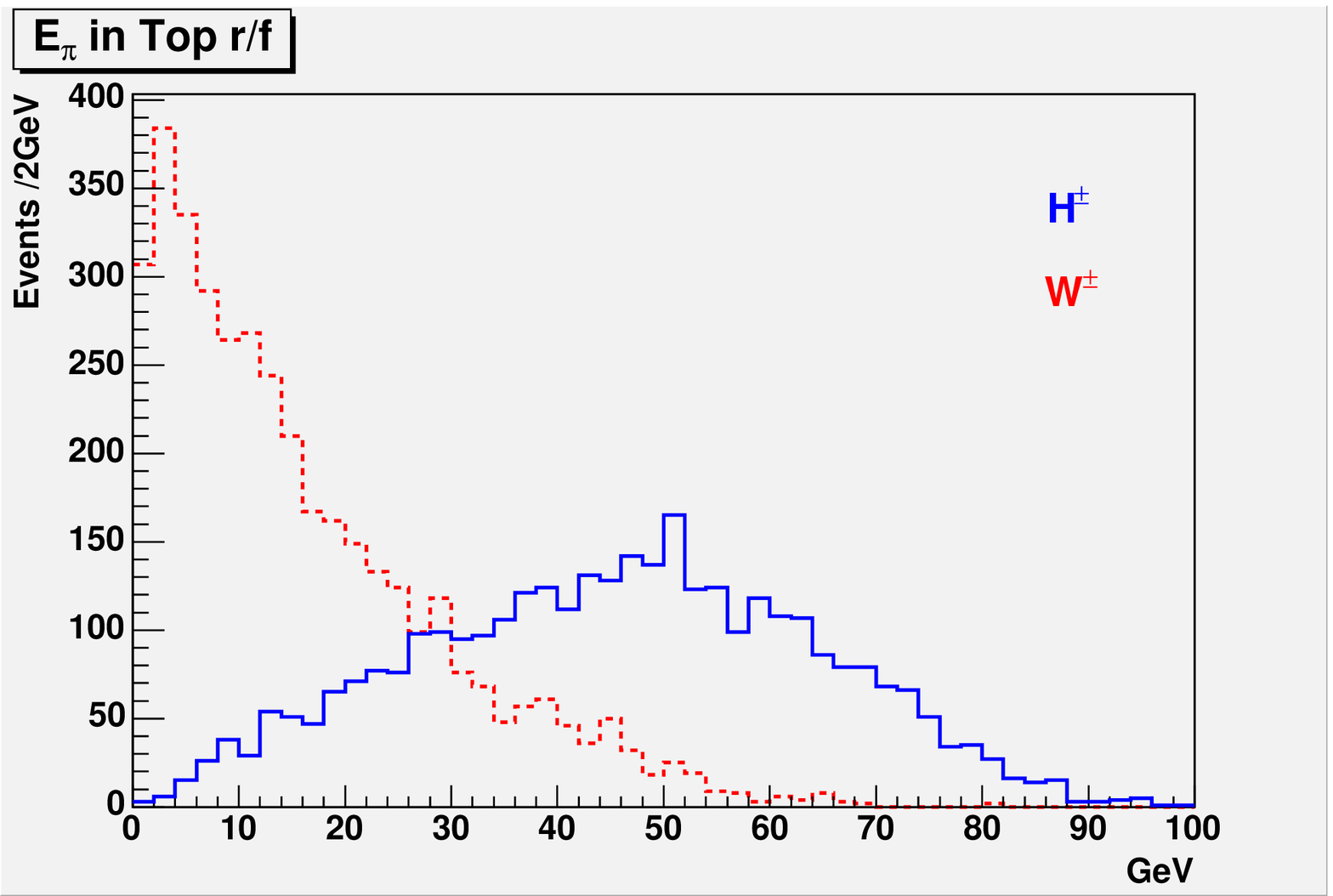}
\includegraphics[width=75mm]{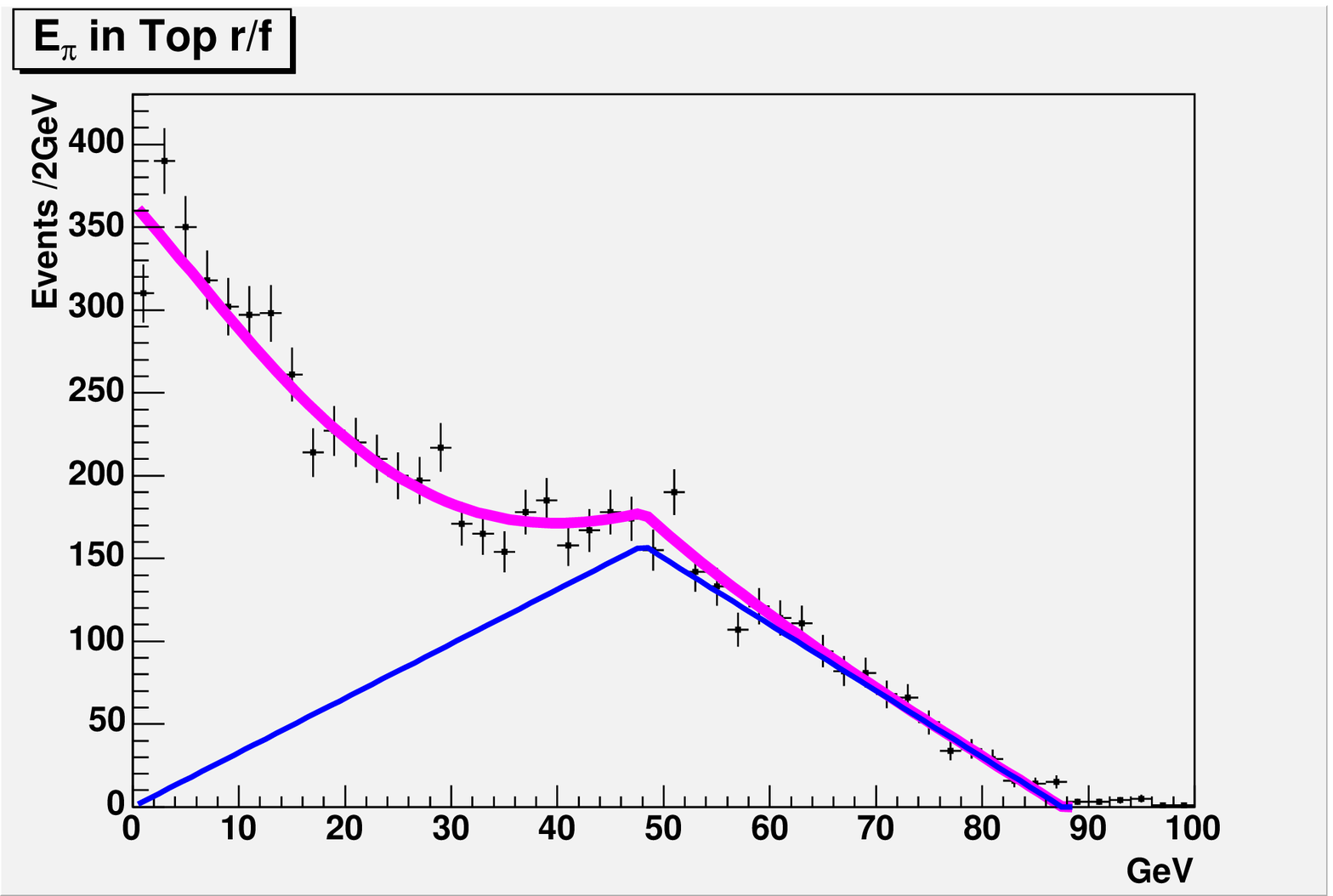}
\includegraphics[width=75mm]{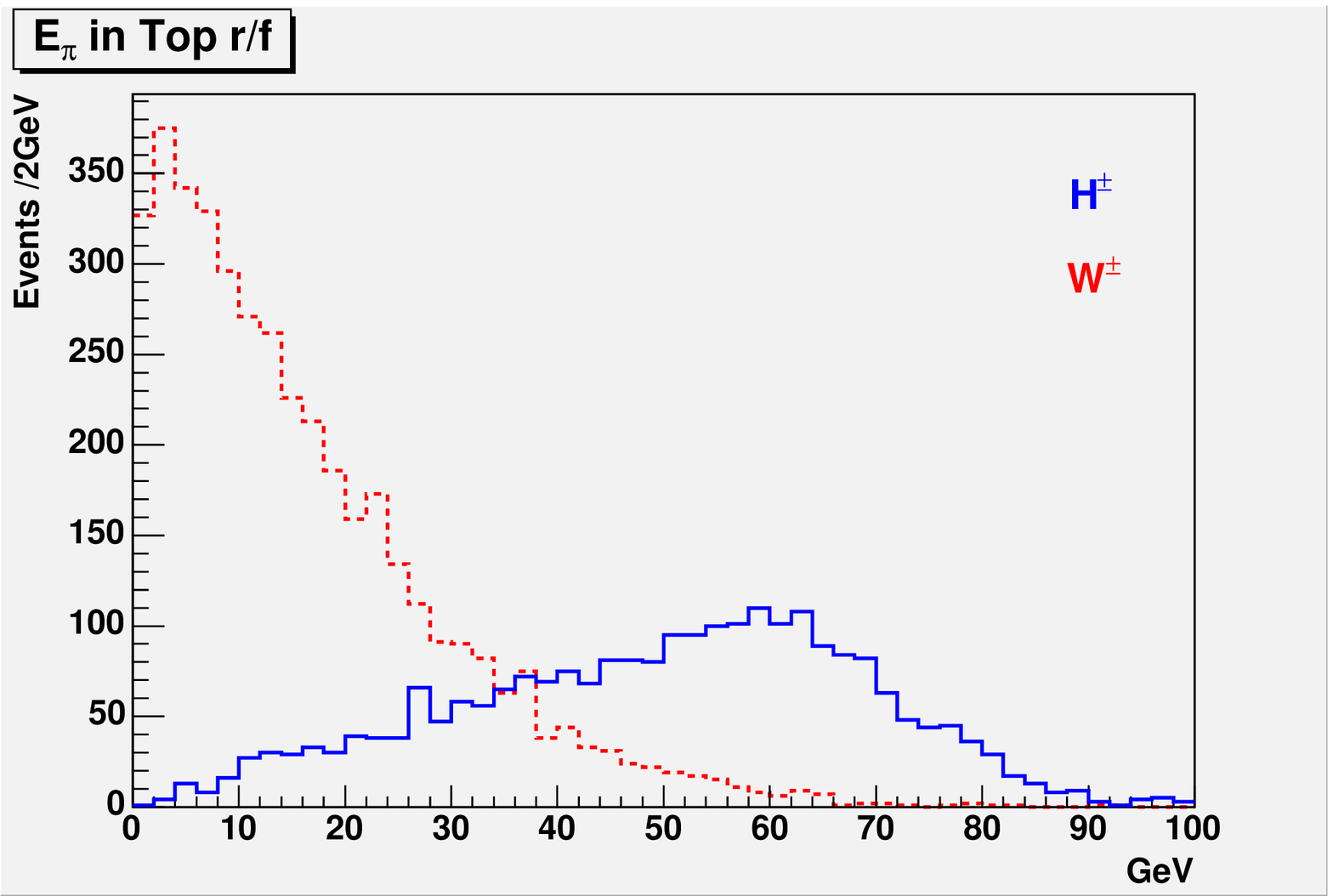}
\includegraphics[width=75mm]{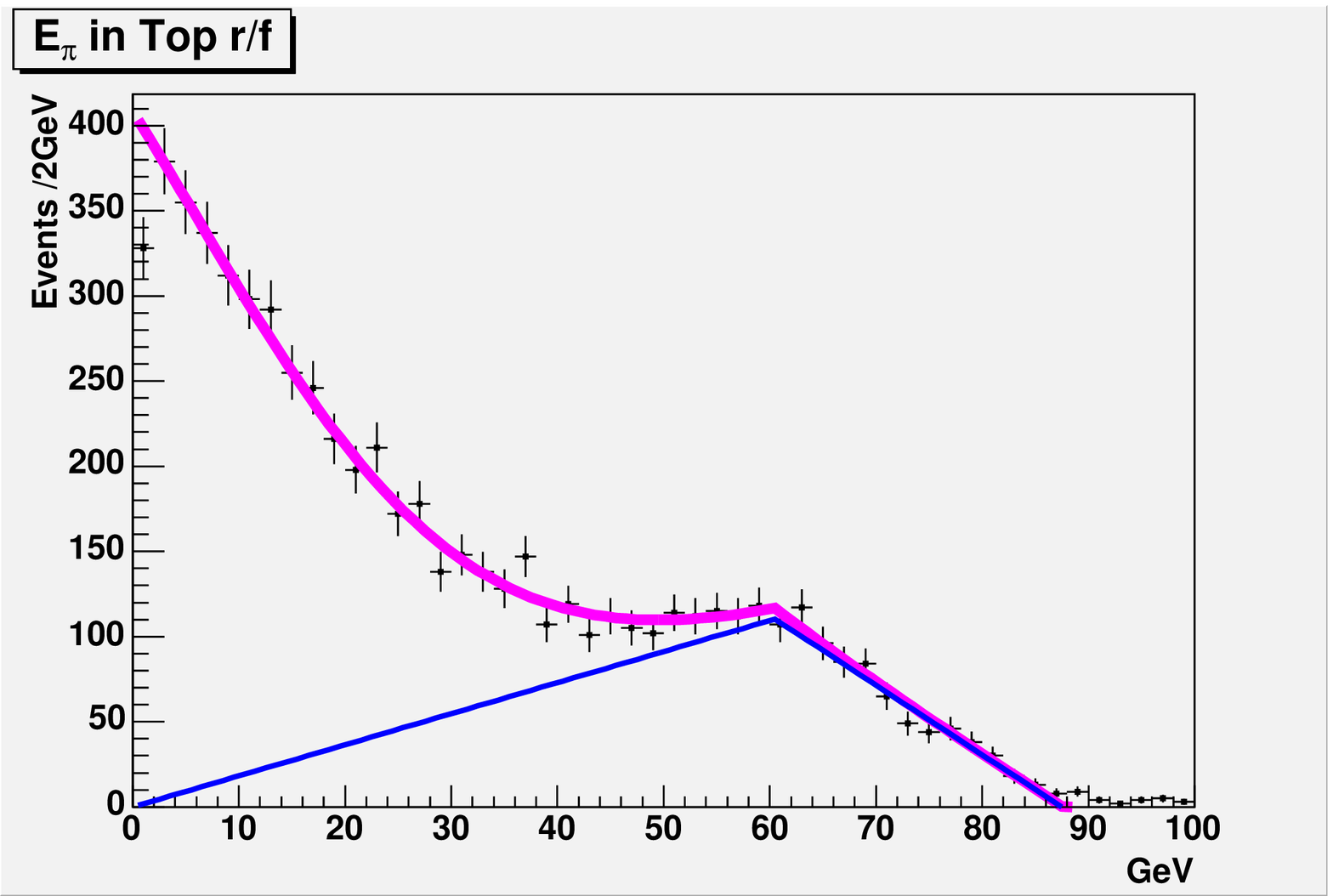}
\caption[.]{Generated $\pi^{\pm}$ energy spectra (left plots)
and the fit (right plots) for the three sets of MSSM parameters
described in the text.  \label{fit}}
\end{figure*}

\begin{center}
\begin{tabular}{rl}
{\em (i)}     & $M_{H^{\pm}}= 129.7 \pm 0.5$~GeV, \cr
{\em (ii)}    & $M_{H^{\pm}}= 129.4 \pm 0.9$~GeV,  \cr
{\em (iii)}   & $M_{H^{\pm}}= 145.5 \pm 0.9$~GeV . \cr
\end{tabular}
\end{center}
Figure~\ref{fit} shows results of the simulations and fits.

\section{CONCLUDING REMARKS}
The tau polarization is a powerful discriminant separating
$H^{\pm}\to\tau^{\pm}\nu$ from $W^{\pm}\to\tau^{\pm}\nu$ 
in top decays in scenarios with rather light charged Higgs
bosons, and with large values of $\tan\beta$, which enhance
the $t b H^{\pm}$  coupling.  In this study, we have 
considered three cases with $M_{H^{\pm}} \approx 130$~GeV
and $\tan\beta = 30$--$50$.  They illustrate how the
$t b H^\pm$ coupling can be enhanced or suppressed
to the extent that $BR(t\to H^{\pm} b)$ varies
from 6\% to 24\%.  A fit to the
energy spectrum of the pion in the $\tau^{\pm}\to\pi^{\pm}\nu$
channel allows one to infer $M_{H^{\pm}}$ with an uncertainty
at the level of $0.5$--$1$~GeV.  This study is a theoretical
level analysis, and no systematics or detector effects have
been included.  On the other hand, only one decay mode 
($\tau^\pm\to\pi^\pm\nu$) has been used, and clearly if one 
would use other decay modes (such as the two-dimensional decay
distributions of the $\tau^\pm\to\rho^\pm\nu$ and
$a_1^\pm\nu$ decays), the sensitivity to $M_{H^{\pm}}$
would improve.  Further work in this direction remains to be done.

\begin{acknowledgments}
The work of E.B. and V.B.
is partly supported by 
RFBR~04-02-16476, RFBR~04-02-17448 and University of 
Russia UR.02.03.028 grants, and Russian Ministry of 
Education and Science NS.1685.2003.2.
Work at ANL is supported in part by the US DOE, Div.
of HEP, Contract W-31-109-ENG-38.  Fermilab is operated by
Universities Research Association Inc.  under contract no.
DE-AC02-76CH02000
with the DOE.
E.B. is grateful to the Fermilab Theoretical Physics 
Department for the kind hospitality
and LCWS05 Organizing Committee for a support. 
\end{acknowledgments}

\end{document}